\documentclass[epjST]{svjour}

\usepackage{graphicx}
\usepackage{graphics}
\usepackage{amssymb}
\usepackage{amsmath}
\usepackage{braket}
\usepackage{epstopdf}

\begin{document}
        
\title{Lattice Supersymmetry and Holography}
\author{Anosh Joseph\inst{1} \fnmsep\thanks{\email{anoshjoseph@iisermohali.ac.in}}}

\institute{\textit{Department of Physical Sciences, \\Indian Institute of Science Education and Research (IISER) - Mohali, \\Knowledge City, Sector 81, SAS Nagar, Punjab 140306, India}}

\abstract{Over the last twenty years, work based on lattice supersymmetry has generated many new results and insights into the non-perturbative nature of string theory, quantum black holes, and gravity. This endeavor is a broad research program encompassing lattice field theory, supersymmetry, string theory, and quantum gravity. In this volume, we look at a selected subset of the topics covering recent progress in lattice supersymmetry and holography.}

\maketitle

\section{Introduction}

Supersymmetric theories play a prominent role in our efforts to understand string theory, strong dynamics, quantum gravity, and various extensions to the Standard Model of particle physics. The holographic duality conjecture reveals the connections between theories of quantum gravity living on curved spacetimes and quantum field theories without gravity living on the boundaries of such spacetimes. This conjecture also provides promising directions for studying the nature of quantum gravity and black holes. 

It is possible to describe certain black hole geometries in terms of the world-volume theories of the D-branes that compose them. These are the maximally supersymmetric Yang-Mills theories in various spacetime dimensions, with many colors, taken in the `t Hooft limit and at finite temperatures. These theories can be strongly coupled in the regimes in which they describe string theory backgrounds, including black holes. Solving these field theories at strong coupling and finite temperature would allow us to directly study the quantum properties of the dual black holes, including their thermodynamic features. As the `t Hooft coupling and the number of colors are reduced, classical and quantum string corrections become more prominent. These less understood limits can be investigated using a lattice formulation - a first-principles definition of quantum field theories. Simulations of lattice discretized field theories can validate the holographic duality conjecture and provide new insight into the non-perturbative structure of string theory and quantum gravity. 

We hope that this special issue would be beneficial to beginning researchers and practitioners in string theory, quantum field theory, and lattice field theory who would want to contribute to this exciting interdisciplinary topic of lattice supersymmetry and its usefulness in testing and validating the holographic duality conjecture. 

\section{A Brief Overview of Contributions}

Below we provide a brief overview of the contributions within this special issue.

In Ref. \cite{Schaich:2022xgy}, David Schaich reviews the recent progress and near future prospects in lattice investigations of supersymmetric field theories and some of the challenges that remain to be overcome. He focuses on the progress in three areas: supersymmetric Yang-Mills (SYM) theories in fewer than four spacetime dimensions, four-dimensional ${\cal N} = 1$ SYM theory and maximally supersymmetric Yang-Mills theory in four dimensions. He also highlights supersymmetric QCD (SQCD) and the sign problem as significant challenges that will be important to address in future work.

In Ref. \cite{Asano:2022jag}, Yuhma Asano outlines some basics of the matrix model conjecture and the gauge/gravity duality conjecture for the matrix models. He reviews various numerical evidence provided for the gauge/gravity duality conjecture for the BFSS and BMN matrix models and their flavored cousins, the Berkooz-Douglas (BD) and Kim-Lee-Yi (KLY) matrix models.

Masanori Hanada and Hiromasa Watanabe review the basic properties of partial deconfinement in Ref. \cite{Hanada:2022wcq} and discuss its applications. The confinement deconfinement transition in gauge theory plays an important role in physics, including describing thermal phase transitions in the dual gravitational theory. In the scenario of partial deconfinement, there is an intermediate phase in which the color degrees of freedom split into the confined and deconfined sectors. The partially deconfined phase is dual to the small black hole that lies between the large black hole and graviton gas. A better understanding of partial deconfinement may explain how gravity emerges from the degrees of freedom of the field theory.

In Ref. \cite{Bliard:2022}, Gabriel Bliard, Ilaria Costa, and Valentina Forini review the lattice study of the Green-Schwarz gauge-fixed string action describing the worldsheet fluctuations about the minimal surface holographically dual to the null cusp Wilson loop. A numerical study of this system using the Monte Carlo method helps evaluate the cusp anomaly of ${\cal N} = 4$ super Yang-Mills. They comment on the discretization, numerical explorations, and challenges for the non-perturbative study of this benchmark model of gauge-fixed worldsheet actions.

In Ref. \cite{Jha:2022}, Raghav Jha proposes additional tests of holography by studying supersymmetric Wilson loops in $p+1$-dimensional maximally supersymmetric Yang-Mills (SYM) theories on a lattice, taken in the large-$N$ limit. In the dual gravity description, this computation involves calculating the area of a fundamental string worldsheet in certain Type II supergravity backgrounds. Even though thermodynamic observables have been computed on the lattice using Monte Carlo methods, and the results agree with the supergravity results in various dimensions, more needs to be done for the gauge-invariant operators, such as the Wilson loop. Jha provides analytical predictions for these loops for various non-conformal D$p$-brane background cases, with $p < 3$, in the large $N$ limit. He also comments on how these can be computed on non-orthogonal lattices for various supersymmetric models.

Daisuke Kadoh and Naoya Ukita in Ref. \cite{Kadoh:2022} propose a supersymmetric gradient flow for four-dimensional ${\cal N} = 1$ SQCD. They gave expressions for the flow equation in the superfield formalism and the component fields formalism in the Wess-Zumino gauge. They also discuss a simplified flow using the gradient of supersymmetric Yang-Mills (SYM) action instead of SQCD action to define a gauge multiplet flow.

\section{Future Research Directions}

In this section, we briefly outline the various research directions the effort of lattice supersymmetry and holography can take in the near future.

\begin{itemize}

\item For the case of the ${\cal N} = 4$ Yang-Mills in four dimensions, there is a famous holographic prediction for the Coulomb coefficient $C(\lambda)$ that it is proportional to $\sqrt{\lambda}$ up to ${\cal O}(1/\sqrt{\lambda})$ corrections. For the $N = \infty$ planar limit more general analytic results have also been obtained. It would be interesting to search for this behavior in more detail, although some promising preliminary results have already been reported.

\item A more detailed understanding of the non-trivial scaling dimension of the simplest conformal primary operator of the four-dimensional ${\cal N} = 4$ SYM theory, the Konishi operator is much needed. Preliminary lattice results have already been obtained and are consistent with existing perturbation theory results.

\item It would be interesting to study the behavior of the four-dimensional ${\cal N} = 4$ SYM theory around the $S$-dual point, where the `t Hooft coupling takes the form $\lambda_{\rm sd} = 4 \pi N$.

\item Another direction is to adjust the scalar potential to study the four-dimensional ${\cal N} = 4$ SYM on the Coulomb branch of the moduli space. In this context, the $S$-duality connects the masses of the U(1)-charged elementary `$W$ bosons' and the magnetically charged topological `t Hooft-Polyakov monopoles. They can be accessed from lattice calculations with appropriate boundary conditions.

\item Non-perturbative lattice calculations can be used to study the free energy of four-dimensional ${\cal N} = 4$ SYM theory. The weak-coupling perturbative prediction and the strong coupling holographic calculation differ by a famous factor of $3/4$. A lattice setup can be used to interpolate between these two coupling regimes. 

\item In the case of maximally supersymmetric Yang-Mills in three dimensions, one can explore, through lattice investigations, the phase transition between the `D2 phase' and the spatially deconfined `D0 phase' dual to a localized black hole geometry.

\item A more detailed and careful non-perturbative investigation of the finite temperature phase diagrams in two-dimensional SYM theories is still needed. These theories also possess rich zero-temperature dynamics, such as the `meson' spectrum and spontaneous symmetry breaking, that are important to explore non-perturbatively.

\item In the context of the BMN model, the critical temperature of the confinement transition can be predicted by perturbative calculations in the weak-coupling regime and by a dual supergravity calculation for strong coupling. An open research direction would be non-perturbatively connecting the transition temperatures by mapping out the intermediate coupling regime, where perturbative and holographic approaches are unreliable.

\item There has been excellent progress in validating and testing holography in the context of BFSS and BMN matrix models. It would be interesting to extend these investigations into more exotic models such as the KLY matrix model.
 
\item Partial deconfinement can be understood as the coexisting phenomenon in the space of the color degrees of freedom. Historically, it was proposed for the SYM theory as the dual of the small black hole phase. An open research direction would be to generalize the idea of partial deconfinement to systems with finite $N$ by utilizing their chiral symmetry.
\end{itemize}

We hope this volume will contribute to constructive discussions on advances in lattice supersymmetry and holography and stimulate further studies.

\subsection*{Acknowledgements}

The author was supported in part by the Start-up Research Grant (No. SRG / 2019 / 002035) from the Science and Engineering Research Board (SERB), Government of India, and in part by the Indian Institute of Science Education and Research (IISER) - Mohali.

\subsection*{Data Availability Statement}

No Data associated in the manuscript.


\end{document}